\documentclass[a4paper,times,fleqn]{cas-dc}
\usepackage[numbers,sort&compress]{natbib}
\usepackage{caption}

\usepackage{graphicx}
\usepackage{subcaption}
\graphicspath{{figures/}{images/}}
\usepackage{float}
\floatstyle{plaintop} % captions above for floats restyled below
\restylefloat{table}
\usepackage{capt-of} % for \captionof{figure} (non-floating figures)
\usepackage{multirow}
\usepackage{longtable}
\usepackage{lscape}
\usepackage[table,xcdraw,dvipsnames]{xcolor}
\usepackage{array}
\usepackage{tabularx}
\usepackage{booktabs}
\usepackage[export]{adjustbox}
\usepackage{colortbl}
\usepackage{enumitem}
\usepackage{multicol}
\usepackage{subcaption}

\usepackage{tikz}
\usetikzlibrary{shapes.geometric,arrows,arrows.meta,positioning,fit,calc,backgrounds}
\usepackage{pgfplots}
\pgfplotsset{compat=1.18}
\usepackage{pgfplotstable}

\usepackage{algorithm}
\usepackage{algpseudocode}

\usepackage{amsmath}
\usepackage{amssymb}

\AtBeginDocument{\hypersetup{colorlinks=true,linkcolor=blue,citecolor=blue,urlcolor=blue}}

\begin{document}

\let\WriteBookmarks\relax
\def\floatpagepagefraction{1}
\def\textpagefraction{.001}

\shorttitle{Routing Cybersecurity Awareness Training by FFM Personality Trait}
\shortauthors{Okwata and Razzaque}

%% Title
\title[mode=title]{%
Routing Cybersecurity Awareness Training by FFM Personality Trait: A Quasi-Experimental Evaluation}

%% Authors
\author[1]{Glory Okwata}
\ead{okwataglory2050@gmail.com}

\author[1]{Mohammad A. Razzaque}[orcid=0000-0002-5572-057X]
\cormark[1]
\ead{m.razzaque@tees.ac.uk}

\affiliation[1]{%
  organization={School of Computing, Engineering and Digital Technologies, Teesside University},
  country={UK}
}

\cortext[cor1]{Corresponding author: Mohammad A. Razzaque, m.razzaque@tees.ac.uk}

%% ---------------------------------------------------------------
\begin{abstract}
Cybersecurity awareness training has historically adopted a one-size-fits-all approach, despite established individual differences in how users process and retain security information. Personality has been proposed as one axis along which training content might be tailored; yet no prior study had implemented and empirically evaluated a complete personality-conditional system end-to-end. This paper reports the design, implementation, and quasi-experimental evaluation of \emph{TailoredSec}, a mobile cybersecurity awareness application that routes training content based on a user's dominant Five-Factor Model (FFM) personality trait, as measured by the ten-item Big Five Inventory (BFI-10). Seventy-four UK-based adults were allocated to a traditional video-training condition ($n = 40$) or a personality-conditional condition ($n = 34$). Both groups completed a four-item scenario-based pre-assessment (scored 0--40), a single training session, and an equivalent post-assessment. The personality-conditional group additionally completed the BFI-10 (Big Five Inventory-10) and was routed to one of four training modules covering five FFM traits (Conscientiousness and Neuroticism share a module). Pre-assessment scores did not differ between groups ($t(69.1) = 0.43$, $p = .67$), confirming baseline equivalence. The personality-conditional group scored significantly higher on the post-assessment ($M = 35.88$, $SD = 5.00$ vs $M = 30.75$, $SD = 10.23$; Welch's $t(58.5) = -2.81$, $p = .003$; Cohen's $d = 0.62$; 95\%,CI $[1.47, 8.79]$ marks), with a pass-rate of 100\% versus 77.5\% (Fisher's exact $p < .01$). A variance ratio of 4.19 indicates a ceiling effect that likely understates the true advantage. Usability was rated 4 or 5 out of 5 by 85.3\% of 68 respondents. These results offer preliminary support for personality-conditional content routing as a feasible design principle for cybersecurity awareness training. Key limitations include non-randomised allocation, an author-developed unvalidated outcome instrument with non-parallel pre/post items, and a single-session design with no retention follow-up. Future work should employ a pre-registered randomised design with a validated cybersecurity awareness scale.
\end{abstract}


\maketitle

%% ---------------------------------------------------------------
\section{Introduction}
\label{sec:intro}

Social engineering attacks, which manipulate human psychology rather than exploit
technical vulnerabilities, have become the dominant entry vector for data
breaches~\cite{langlois2023dbir,mashtalyar2021social}. The 2025 Verizon Data Breach
Investigations Report documented that 60\% of all breaches involved the human element \cite{Humanerr47:online},
with social engineering and pretexting incidents increasing by 50\% year on year. The
proliferation of generative AI tools further amplifies this threat by enabling attackers
to craft personalised, linguistically polished lures at scale~\cite{10198233,aslam2024ai}.
Yet the predominant organisational response---mandatory annual awareness training
delivered identically to all employees---has changed little in its basic
structure~\cite{khando2021enhancing}. The training is still uniform, but the attacks are not.

Uniform training fails for a well-documented reason: standardised content cannot
accommodate the variation in prior knowledge, cognitive style, risk perception, and
motivation that employees bring to a training session~\cite{aldawood2019reviewing}.
People differ, and they differ in how they encode threat information, in their susceptibility
to specific social engineering tactics~\cite{workman2008wisecrackers,bullee2018anatomy},
and in the pedagogical formats that best support knowledge
retention~\cite{kim2016how,elsabagh2021adaptive}. A growing
body of evidence links personality traits, particularly those captured by the Five-Factor
Model (FFM)~\cite{mccrae1992introduction}, to cybersecurity
behaviour~\cite{shappie2019personality,gratian2018correlating,BALTUTTIS2024103741}.
Conscientious individuals exhibit stronger security compliance~\cite{kennison2020taking,kalhoro2022personality},
while extraversion and high agreeableness are associated with greater susceptibility to
phishing and pretexting~\cite{albladi2017personality,halevi2016cultural,lawson2020phishing}.
FFM trait profiles could therefore serve as a routing variable for training content;
yet no prior study has built a complete mobile application around this idea, deployed
it with a real user population, and compared it against a control condition.

Three prior lines of work come closest. Uebelacker and Quiel~\cite{uebelacker2014social}
proposed a conceptual mapping between FFM traits and Cialdini's persuasion
principles~\cite{cialdini2009influence}, suggesting trait-specific coping mechanisms for
security training, but they did not implement or evaluate a training system. Gianotti et
al.~\cite{gianotti2019model} proposed incorporating FFM traits into an educational
recommender system architecture for general learning, without a cybersecurity focus or
empirical evaluation. Thorp et al.~\cite{thorp2023association} demonstrated that FFM
traits moderated the effectiveness of virtual-reality-based cybersecurity training, but
the training content itself was not varied by trait. The present study addresses this
gap by implementing a full system in which BFI-10 scores determine which of four training
modules a participant receives and by evaluating the resulting system against a
traditional-training control.

This paper makes three contributions. \textbf{First}, we present the design and
implementation of \emph{TailoredSec}, a cybersecurity awareness mobile application
that operationalises FFM trait scores as a content-routing variable, with the routing
rule, content mapping (five traits to four modules), tie-breaking, and BFI-10 scoring
algorithm fully specified to a replication standard. \textbf{Second}, we report a
quasi-experimental comparison ($n = 74$; non-randomised; baseline-equivalent on
pre-assessment, $t(69.1)=0.43$, $p=.67$) showing a moderate, statistically significant
post-assessment advantage of personality-conditional training over traditional video
training (Welch's $t(58.5)=-2.81$, $p=.003$, Cohen's $d = 0.62$, 95\,\%\,CI for mean
difference $[1.47, 8.79]$), with 100\% versus 77.5\% pass-rates (Fisher's exact
$p < .01$). \textbf{Third}, we make the design choices, psychometric constraints, and
analytic limitations of the study explicit---including the ceiling effect visible in the
post-test variance ($F_\text{ratio}=4.19$), the individual-routing limitations of the
BFI-10, and the non-parallel pre/post instrument---and we use these to formulate a
pre-registered replication design that the research community can build upon.

The remainder of the paper is structured as follows. Section~\ref{sec:related} reviews
the literature on social engineering susceptibility, FFM-based training differentiation,
and mobile platforms for security education. Section~\ref{sec:method} describes the
study design, participants, ethics, instruments, and analysis plan. Section~\ref{sec:system}
details the system architecture and implementation of \emph{TailoredSec}.
Section~\ref{sec:results} presents the empirical results. Section~\ref{sec:discussion}
interprets the findings, addresses limitations, and proposes future directions.
Section~\ref{sec:conclusion} concludes.

%% ---------------------------------------------------------------
\section{Background and Related Work}
\label{sec:related}

\subsection{Social Engineering as the Primary Human-Layer Threat}

Social engineering exploits the psychological and social vulnerabilities of individuals
rather than technical weaknesses in information systems, making it particularly
difficult to defend against through purely technical
means~\cite{mouton2014social,hadnagy2011social}. Attacks range from phishing and
spear-phishing to pretexting, baiting, tailgating, and vishing, and typically exploit
Cialdini's principles of influence---reciprocity, commitment, social proof, authority,
liking, and scarcity~\cite{cialdini2009influence,bullee2018anatomy}. Across reported incidents, between 70\% and 95\% involve a human element, at a cost to
organisations ranging from thousands to millions of pounds in direct financial loss,
regulatory penalty, and reputational
damage~\cite{langlois2023dbir,zwilling2022cyber,tawalbeh2023factors}. Security awareness training is the most widely deployed form of human-layer protection 
and is at least as important as technical perimeter controls~\cite{khando2021enhancing}.

\subsection{Limitations of Uniform Security Awareness Training}

Conventional security awareness training programmes share a common architecture: a
standardised curriculum delivered via presentation slides, e-learning modules, or
mandatory video content, typically on an annual or biannual schedule. Evaluation of
these programmes reveals a consistent pattern: short-term knowledge gains that dissipate
rapidly, low engagement, negligible impact on long-term behaviour, and high recidivism
on phishing simulations~\cite{aldawood2019reviewing,kim2016how}. Lawson et al.~\cite{lawson2020phishing}
demonstrated that within an organisation subjected to repeated simulated phishing
campaigns, a stable minority of employees continued to click malicious links despite
training, suggesting that a single, uniform intervention cannot address the heterogeneity
of individual risk profiles. Parsons et al.~\cite{parsons2017development}, in developing
the Human Aspects of Information Security Questionnaire (HAIS-Q), similarly found that
security behaviours vary substantially across individuals and domains, reinforcing the
case for differentiated rather than uniform training approaches. At its core, the one-size-fits-all model assumes all learners respond equivalently to
the same content and framing, a claim contradicted by decades of educational psychology
research~\cite{furnham1992personality,komarraju2011bigfive,du2017analysis}.

\subsection{Personality and Cybersecurity Behaviour}

The FFM, also known as the OCEAN model, covers five broad dimensions of human
personality---Openness to experience, Conscientiousness, Extraversion, Agreeableness,
and Neuroticism---and is the dominant framework for personality assessment in academic
and applied settings~\cite{mccrae1992introduction}. Multiple independent studies now
link FFM traits to information security behaviour.
Gratian et al.~\cite{gratian2018correlating} found that Openness and Conscientiousness
were among the strongest individual-level predictors of security intention in a
sample of 500 participants. Shappie et al.~\cite{shappie2019personality} reported that
Conscientiousness and Agreeableness positively predicted self-reported cybersecurity
compliance, while Neuroticism was negatively associated with secure password practices.
Kennison and Chan-Tin~\cite{kennison2020taking} and Kalhoro et
al.~\cite{kalhoro2022personality} identified low Conscientiousness and high Extraversion
as consistent risk factors for security-compromising behaviour in workplace samples.
Albladi and Weir~\cite{albladi2017personality} demonstrated that personality traits
mediated the relationship between online behaviour and victimisation, with trust and
impulsivity as key intervening variables. Across these studies, the FFM---security behaviour link is consistent, with typical
effect sizes ($r$) in the 0.1--0.3 range at the trait level~\cite{halevi2013spear}.

Two caveats deserve mention. \textbf{First}, the trait--susceptibility relationships
observed in the literature are primarily correlational and mediated by variables such as
online exposure and risk perception; personality is not destiny with respect to security
behaviour. \textbf{Second}, the BFI-10 used in the present study was designed and validated for
group-level research. Rammstedt and John~\cite{rammstedt2007a} explicitly caution that
its two-item per-trait structure yields retest reliabilities of $r = .49$ (Neuroticism,
German sample) to $r = .79$ (Extraversion, US sample), which is insufficient for
individual clinical diagnosis. This psychometric limitation of using BFI-10 scores for
individual content-routing decisions is discussed further in Section~\ref{sec:discussion}.

\subsection{FFM Traits, Susceptibility, and Coping Mechanisms}

Uebelacker and Quiel's~\cite{uebelacker2014social} theoretical framework provides the
most direct basis for linking individual FFM traits to training content design. Their
analysis of the personality--persuasion correspondence identified the following
trait-specific patterns and coping recommendations, subsequently extended by
Papatsaroucha et al.~\cite{papatsaroucha2021survey}:

\textit{Openness:} Individuals high in Openness are characterised by intellectual
curiosity and a propensity for novel experiences, which increases susceptibility to
baiting attacks~\cite{alseadoon2012susceptible}. Training formats that provide
interactive, exploratory, and intellectually stimulating content---such as gamification
and problem-based scenarios---are most effective for this
group~\cite{uebelacker2014social,thorp2023association}.

\textit{Conscientiousness:} Conscientious individuals tend to follow rules and respect
authority, which paradoxically increases their susceptibility to authority-based
manipulation~\cite{roberts2012conscientiousness}. General security awareness training
covering policy and procedural norms is recommended, as this format aligns with their
normative disposition~\cite{uebelacker2014social,halevi2016cultural}.

\textit{Extraversion:} Extraverted individuals are socially motivated and enjoy
interactive media; they are more vulnerable to social engineering via social media and
communication platforms~\cite{albladi2017personality}. Personalised, rewarding, and
socially framed content is most effective~\cite{uebelacker2014social}.

\textit{Agreeableness:} High Agreeableness is associated with trust, compliance, and
prosocial motivation, increasing vulnerability to authority, reciprocity, and liking
appeals~\cite{papatsaroucha2021survey}. Narrative and storytelling-based training, presenting social engineering through concrete
case studies, has been recommended to recalibrate overgeneralised
trust~\cite{uebelacker2014social}.

\textit{Neuroticism:} Neurotic individuals tend to exhibit anxiety and risk-aversion, which
can paradoxically limit susceptibility by heightening vigilance toward suspicious
communications~\cite{bansal2010impact}. General security awareness training, similar to
that for Conscientiousness, is recommended given the limited incremental vulnerability
of this trait~\cite{uebelacker2014social}.

These recommendations had not, however, been implemented and tested in a deployed
training system before the present study. Table~\ref{tab:ffm-summary} summarises the
trait characteristics and corresponding module formats.

\begin{table}[!t]
\caption{Summary of FFM personality traits, cybersecurity susceptibility, and
  training recommendations. Susceptibility levels adapted from
  Uebelacker and Quiel~\protect\cite{uebelacker2014social} and
  Papatsaroucha et al.~\protect\cite{papatsaroucha2021survey}.}
\label{tab:ffm-summary}
\footnotesize
\begin{tabular}{@{}p{2.2cm}p{2.4cm}p{2.4cm}@{}}
\toprule
\textbf{Trait} & \textbf{Susceptibility vector} & \textbf{Recommended format} \\
\midrule
\rowcolor[HTML]{DEEAF6}
Openness & Baiting, curiosity exploitation & Interactive / gamified content \\
Conscientiousness & Authority, commitment appeals & General awareness video \\
\rowcolor[HTML]{DEEAF6}
Extraversion & Phishing, social-media manipulation & Reward-based / personalised content \\
Agreeableness & Reciprocity, liking, social proof & Storytelling / narrative video \\
\rowcolor[HTML]{DEEAF6}
Neuroticism & Lower susceptibility via vigilance & General awareness video \\
\bottomrule
\end{tabular}
\end{table}

Figure~\ref{fig:routing} illustrates the resulting content routing structure, showing
the 5-to-4 module collapse.

\begin{center}
\begin{tikzpicture}[
  traitbox/.style={rectangle, draw, rounded corners=2pt, fill=blue!15,
    text width=2.0cm, align=center, font=\footnotesize, minimum height=0.65cm},
  modbox/.style={rectangle, draw, rounded corners=2pt, fill=orange!20,
    text width=2.8cm, align=center, font=\footnotesize, minimum height=0.65cm},
  arr/.style={-{Stealth[length=2.5mm]}, thick, blue!70}
]
  \node[font=\small\bfseries] (lbl1) at (0,0) {BFI-10 Dominant Trait};
  \node[font=\small\bfseries] (lbl2) at (5.5,0) {Training Module};

  \node[traitbox] (O)  at (0,-0.9)  {Openness};
  \node[traitbox] (E)  at (0,-1.7)  {Extraversion};
  \node[traitbox] (A)  at (0,-2.5)  {Agreeableness};
  \node[traitbox] (C)  at (0,-3.3)  {Conscientiousness};
  \node[traitbox] (Ne) at (0,-4.1)  {Neuroticism};

  \node[modbox] (M1) at (5.5,-0.9)  {Swipeable Cards\\\footnotesize(Baiting/Phishing)};
  \node[modbox] (M2) at (5.5,-1.7)  {Audio Podcast\\\footnotesize(Reward-based)};
  \node[modbox] (M3) at (5.5,-2.5)  {Storytelling Video\\\footnotesize(Case-study narrative)};
  \node[modbox] (M4) at (5.5,-3.7)  {General Video\\\footnotesize(Security awareness)};

  \draw[arr] (O.east)  -- (M1.west);
  \draw[arr] (E.east)  -- (M2.west);
  \draw[arr] (A.east)  -- (M3.west);
  \draw[arr] (C.east)  -- (M4.west);
  \draw[arr] (Ne.east) -- (M4.west);

  %% Brace for collapse
  \draw[decorate,decoration={brace,amplitude=4pt},thick,gray]
    ([xshift=0.05cm]C.north east) -- ([xshift=0.05cm]Ne.south east)
    node[midway, right=0.1cm, font=\tiny, gray] {shared};
\end{tikzpicture}

\captionof{figure}{Content routing structure. BFI-10 dominant trait scores map five FFM traits
  onto four training modules. Conscientiousness and Neuroticism share a general
  security awareness video module.}
\label{fig:routing}
\end{center}

\subsection{Adaptive and Personalised Learning in Security Training}

Personalised e-learning systems modify content based on learner characteristics to
improve engagement and outcomes~\cite{elsabagh2021adaptive}. In general education, personality-informed personalisation has clear empirical backing:
Komarraju et al.~\cite{komarraju2011bigfive} found FFM traits explained 14\% of GPA
variance, and Lai et al.~\cite{lai2020automatic} showed that personality profiles inferred
from online learning behaviour could reliably guide content selection.
In the cybersecurity domain, however, such systems remain rare. Giannakas et
al.~\cite{giannakas2015cyberaware} developed a game-based mobile application for
cybersecurity education aimed at younger learners, but did not personalise content by
personality trait. Sudha et al.~\cite{sudha2023impact} demonstrated that smartphone-based
interactive modules improved cybersecurity knowledge at the high-school level, again
without trait-based personalisation. No prior cybersecurity study had deployed and evaluated a complete trait-routing system.
The present study fills that gap by deploying a mobile application that routes
cybersecurity awareness content by FFM trait and comparing it against a control condition.

One terminological point: the system implements \emph{personality-conditional content
routing}---a one-shot assignment based on a personality profile---not \emph{adaptive
learning} in the technical sense, which requires continuous updating of a learner model
from real-time performance data (as in intelligent tutoring systems or knowledge-tracing
models)~\cite{elsabagh2021adaptive}. This distinction is maintained throughout.

%% ---------------------------------------------------------------
\section{Methodology}
\label{sec:method}

\subsection{Study Design}

This study employed a quasi-experimental between-subjects design with pre- and
post-assessment measurement. Participants were allocated to one of two conditions: a
traditional training condition (Sample~1, $n = 40$) and a personality-conditional
training condition (Sample~2, $n = 34$). The unit of analysis was the individual
participant. The study was not randomised; participants who contacted the researcher
first were allocated to Sample~1 until the target for that group was reached, after
which subsequent participants were allocated to Sample~2. This allocation mechanism
produced groups that did not differ on pre-assessment performance (see
Section~\ref{sec:results:baseline}), but it limits causal inference. The study is therefore
described throughout as quasi-experimental.

\subsection{Participants and Recruitment}

A convenience sample of 112 UK-based adults was approached via direct researcher contact
and participant referral networks between October and December 2023. Inclusion criteria
were: (i) UK residence, (ii) ownership of an Android smartphone capable of sideloading
an unsigned APK, and (iii) availability for a single self-directed session of
approximately 15--30 minutes. No exclusion criteria related to prior cybersecurity
knowledge were applied. Seventy-four participants completed the study, yielding a
completion rate of 66.1\%. The 38 non-completers ($33.9\%$) did not engage with the
application; the most plausible reason is reluctance to enable the installation of an
unsigned APK, which requires non-default device settings. This attrition mechanism
introduces a selection bias towards participants who are more technically literate or
security-confident, which is discussed in Section~\ref{sec:limitations}.

\begin{figure*}[ht]
\centering
\begin{tikzpicture}[
  box/.style={rectangle, draw=black, rounded corners=3pt, text width=4.8cm,
              align=center, minimum height=0.9cm, font=\small},
  attrbox/.style={rectangle, draw=gray, rounded corners=3pt, text width=3.8cm,
              align=center, minimum height=0.75cm, font=\footnotesize, fill=gray!10},
  arr/.style={-{Stealth[length=3mm]}, thick},
  side/.style={-{Stealth[length=2.5mm]}, thick, gray}
]
  \node[box, fill=blue!10] (enrol)
    {Assessed for eligibility\\\textbf{$n = 112$}};

  \node[attrbox, right=1.2cm of enrol] (excl)
    {Did not complete ($n = 38$):\\ APK sideload barriers\\ or no response};

  \node[box, below=0.7cm of enrol, fill=blue!10] (rand)
    {Enrolled and allocated ($n = 74$)};

  \node[box, below left=0.7cm and 1.2cm of rand, fill=orange!15] (s1)
    {Sample~1 --- Traditional\\Allocated: $n = 40$};

  \node[box, below right=0.7cm and 1.2cm of rand, fill=green!10] (s2)
    {Sample~2 --- Personality-conditional\\Allocated: $n = 34$};

  \node[box, below=0.7cm of s1, fill=orange!10] (s1c)
    {Completed training ($n = 40$)};

  \node[box, below=0.7cm of s2, fill=green!8] (s2c)
    {Completed training ($n = 34$);\\33 received personality-routed\\content; 1 passed pre-assessment};

  \node[box, below=0.7cm of s1c, fill=orange!20] (s1a)
    {Analysed ($n = 40$)};

  \node[box, below=0.7cm of s2c, fill=green!20] (s2a)
    {Analysed ($n = 34$)};

  \draw[arr] (enrol) -- (rand);
  \draw[side] (enrol.east) -- (excl.west);
  \draw[arr] (rand.south) -- ++(0,-0.35) -| (s1.north);
  \draw[arr] (rand.south) -- ++(0,-0.35) -| (s2.north);
  \draw[arr] (s1) -- (s1c);
  \draw[arr] (s2) -- (s2c);
  \draw[arr] (s1c) -- (s1a);
  \draw[arr] (s2c) -- (s2a);
\end{tikzpicture}
\caption{CONSORT-style participant flow diagram. Of 112 individuals approached, 74 completed
  the study. Non-completers ($n=38$) did not install the application, most likely due to
  reluctance to enable sideloading of an unsigned APK. One participant in Sample~2 passed the
  pre-assessment and proceeded directly to post-assessment without receiving personality-routed
  training content.}
\label{fig:consort}
\end{figure*}
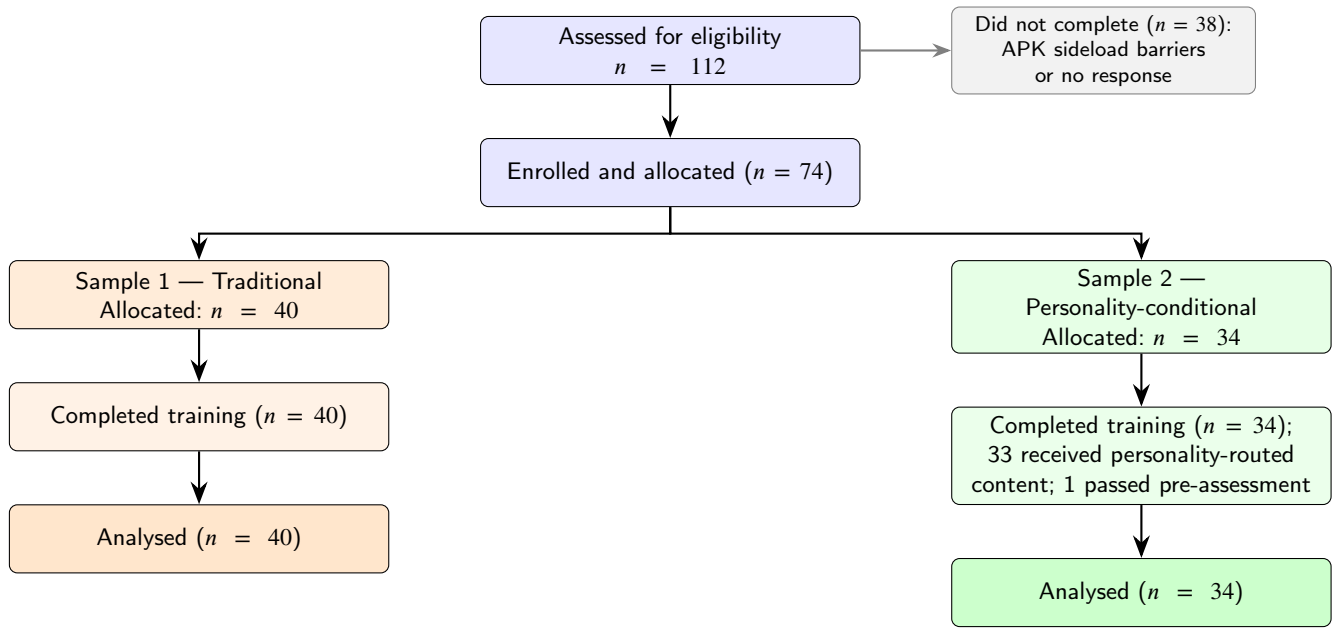

The CONSORT-style (Consolidated Standards of Reporting Trials) participant flow diagram is presented in Figure~\ref{fig:consort}.

\subsection{Ethics}

The study was conducted under the ethical framework of Teesside University's School of
Computing, Engineering, and Digital Technologies. No personally identifying information
(PII) was collected at any stage. Participants were informed of the study's purpose,
voluntary nature, data handling procedures, and their right to withdraw at any point
without consequence via a consent notification displayed on the application's onboarding
screen. All data were stored on Google Firebase Firestore under enforced security rules,
accessible only to the research team. Quiz responses were recorded anonymously using a
session identifier that could not be linked to any individual outside the session.

\subsection{Instruments}

\subsubsection{Personality Assessment: BFI-10}

Personality was assessed using the Big Five Inventory-10 (BFI-10)~\cite{rammstedt2007a},
a validated 10-item instrument measuring each of the five FFM dimensions with two items
per trait on a five-point Likert scale (1 = Disagree strongly, 5 = Agree strongly). The
BFI-10 was selected over the full BFI-44~\cite{john1991bigfive} or the NEO-PI-R
(60 items) on the grounds of brevity and acceptability within a mobile application context. It captures approximately 70\% of the variance of the BFI-44 and has demonstrated
acceptable construct validity~\cite{rammstedt2007a,rammstedt2014limits}. Items 1, 3, 4,
5, and 7 are reverse-scored prior to summing (see Section~\ref{subsec:bfi10scoring}).

\subsubsection{Social Engineering Awareness: Scenario-Based MCQ}

We measured social engineering awareness by developing a four-item
, scenario-based multiple-choice questionnaire administered at both pre- and post-assessment. Each item presented a realistic social engineering scenario drawn from
five threat categories (phishing, vishing, tailgating, pretexting, and baiting), with
four response options scored 0 (no credit) or 10 (full credit), yielding a total score
of 0--40 per assessment. A passing score was defined a priori as $\geq 30$, equivalent
to correctly answering at least three of the four scenarios.

% Important limitations of this instrument must be noted. First, the pre- and
% post-assessment items were intentionally non-parallel (different scenarios were used to
% reduce testing-effect confounds while covering the same threat categories); this
% non-parallelism means that the pre/post comparison is not strictly on a stable
% measurement scale. Second, the instrument was not validated against established scales
% such as the HAIS-Q~\cite{parsons2017development} or the SeBIS, and no item-level
% reliability analysis (e.g., Cronbach's $\alpha$) was conducted. Third, with only four
% items, the assessment has near-zero discriminating power at the upper end of the ability
% distribution after effective training (the ceiling effect documented in
% Section~\ref{sec:results:primary}). Future replication studies should use a validated,
% multi-item cybersecurity knowledge instrument.

\subsubsection{Usability Feedback: Five-Item Likert Survey}

Post-training usability was assessed via an anonymous five-point Likert survey
administered through Google Forms, covering four dimensions: overall usability,
perceived quality of adaptive content matching, helpfulness of app features for
understanding social engineering, and ease of use.

\subsection{Construct Operationalisation}

Table~\ref{tab:constructs} provides a complete operationalisation of all constructs
measured in the study.

\begin{table}[H]
\caption{Construct operationalisation table.}
\label{tab:constructs}
\footnotesize
\begin{tabular}{@{}p{1.8cm}p{1.9cm}p{2.0cm}p{1.8cm}@{}}
\toprule
\textbf{Construct} & \textbf{Instrument} & \textbf{Items / range} & \textbf{Source} \\
\midrule
\rowcolor[HTML]{DEEAF6}
Personality trait & BFI-10 & 10 items; 5-pt Likert; 2 items/trait; reverse-scored & Rammstedt \& John \cite{rammstedt2007a} \\
SE awareness & Scenario MCQ & 4 items; 0--40\,pts; non-parallel pre/post & Author-developed \\
\rowcolor[HTML]{DEEAF6}
Usability & Likert survey & 4 dimensions; 5-pt scale & Author-developed \\
App performance & Apptim & CPU, memory, energy, threads & Apptim \cite{ahmad2023analysis} \\
\bottomrule
\end{tabular}
\end{table}

\subsection{Adaptation Decision Rule}
\label{subsec:adaptationrule}

The formal specification of the personality-conditional content routing logic is given
in Algorithm~\ref{alg:routing}. After scoring the BFI-10, the dominant trait is
determined as the argmax over the five trait scores. In the case of a tie between two or
more traits, the system applies a predefined priority order (Openness $>$ Agreeableness
$>$ Extraversion $>$ Conscientiousness $>$ Neuroticism) that allocates ties to the
trait with the most differentiated training module. Conscientiousness and Neuroticism
share the general awareness video module, so the 5-trait input space maps onto 4 training
modules (see Figure~\ref{alg:routing}). Participants who scored $\geq 30$ on the pre-assessment (passing) were shown a
pass screen and could optionally proceed to the training and post-assessment; those who
failed ($< 30$) proceeded directly to the BFI-10 and the routed training module.

\begin{algorithm}[H]
\caption{BFI-10 Personality-Conditional Content Routing}
\label{alg:routing}
\begin{algorithmic}[1]
\Require BFI-10 responses $r_1, \ldots, r_{10} \in \{1,2,3,4,5\}$
\Ensure Training module $M \in \{\text{Cards}, \text{Video}, \text{Podcast}, \text{Story}\}$

\State \textbf{Reverse-score} items 1,3,4,5,7: $r_i \leftarrow 6 - r_i$

\State \textbf{Compute trait scores:}
\State $E \leftarrow r_1 + r_6$ \Comment{Extraversion}
\State $A \leftarrow r_2 + r_7$ \Comment{Agreeableness}
\State $C \leftarrow r_3 + r_8$ \Comment{Conscientiousness}
\State $N \leftarrow r_4 + r_9$ \Comment{Neuroticism}
\State $O \leftarrow r_5 + r_{10}$ \Comment{Openness}

\State $\text{dominant} \leftarrow \arg\max_{\tau \in \{O,C,E,A,N\}} \tau$ \quad (ties: priority $O{>}A{>}E{>}C{>}N$)

\If{dominant $= O$}   \Return $M \leftarrow \text{Swipeable Cards}$ \EndIf
\If{dominant $\in \{C, N\}$} \Return $M \leftarrow \text{General Awareness Video}$ \EndIf
\If{dominant $= E$}   \Return $M \leftarrow \text{Audio Podcast}$ \EndIf
\If{dominant $= A$}   \Return $M \leftarrow \text{Storytelling Video}$ \EndIf
\end{algorithmic}
\end{algorithm}

\noindent

\subsection{Procedure}

Participants received an invitation message containing installation instructions for the
\emph{TailoredSec} Android APK and a participation information sheet. The session
was self-directed and asynchronous; participants completed all elements (pre-assessment,
training, post-assessment, and optional feedback survey) in a single continuous session,
estimated to take 15--30 minutes. The onboarding screen displayed the consent statement.
Participants allocated to Sample~1 received a general cybersecurity awareness video
followed by the post-assessment, without completing the BFI-10. Participants allocated
to Sample~2 who failed the pre-assessment ($< 30$ marks) completed the BFI-10, received
the routed training module, and then completed the post-assessment.

\subsection{Analysis Plan}

The primary outcome variable was the post-assessment score (0--40). The primary
hypothesis was that the personality-conditional group would score higher on the
post-assessment than the traditional group (directional hypothesis; one-tailed test).
Pre-assessment scores were analysed using Welch's independent-samples $t$-test to
assess baseline equivalence. The primary outcome comparison used Welch's
independent-samples $t$-test (chosen over Student's $t$-test to account for unequal
sample variances). Effect size was reported as Cohen's $d$, computed from pooled standard
deviation, with 95\% confidence intervals on the mean difference derived from Welch's
standard error. The pass-rate contingency (Fisher's exact test was used, rather than
Pearson's $\chi^2$, because one cell had an expected count of zero) was computed for the
$\geq 30$-mark threshold. Variance homogeneity was examined via the variance ratio
$s^2_{\text{Trad}} / s^2_{\text{PC}}$; a ratio substantially greater than 1 was
interpreted as evidence of a ceiling effect in the personality-conditional group.

%% ---------------------------------------------------------------
\section{System Design and Implementation}
\label{sec:system}

\subsection{Technology Stack and Architecture}

\begin{figure*}
\centering
\includegraphics[width=0.75\textwidth]{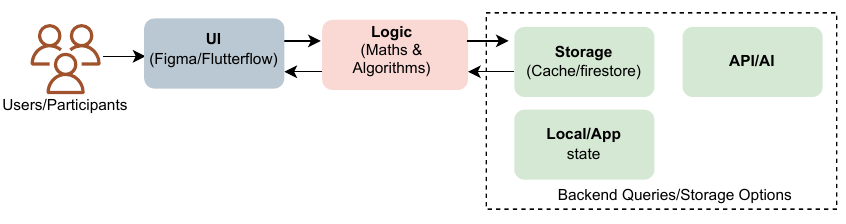}
\caption{Overview of \emph{TailoredSec} system architecture. The Android client
  manages application state in-memory (FlutterFlow AppState cache); scores and session
  data are persisted to Firebase Firestore on completion of each quiz segment.}
\label{fig:arch}
\end{figure*}

The \emph{TailoredSec} application was implemented as a native Android application
using FlutterFlow~\cite{FlutterF10:online}, a visual no-code/low-code development
environment built on the Flutter framework, which enables cross-platform deployment from
a single codebase. Firebase Firestore~\cite{Firestor52:online} was used as the
serverless NoSQL backend for real-time data persistence, providing schema-flexible
storage for pre- and post-assessment scores and session metadata. The user interface was
prototyped in Figma~\cite{hidayanti2023uiux} prior to implementation. No user
authentication was implemented, consistent with the ethical requirement for anonymous
participation; a session-level identifier was generated at runtime and discarded upon
session completion. The high-level system architecture, illustrating the data flow
between the client (Android device), app state cache, and Firebase Firestore backend, is
shown in Figure~\ref{fig:arch}.

\subsection{Application Workflow}

\begin{figure}
\centering
\includegraphics[width=0.46\textwidth]{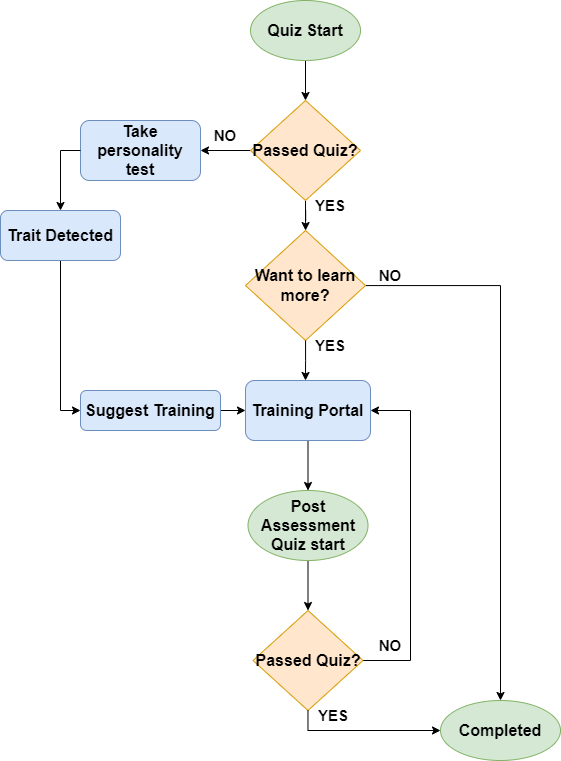}
\caption{Application logical flow. Participants who pass the pre-assessment
  ($\geq 30$) may exit or proceed directly to post-assessment. Participants who
  fail are routed through the BFI-10 (personality-conditional group) or directly to
  standard training (traditional group), and then to post-assessment.}
\label{fig:appflow}
\end{figure}

The application implements a conditional branching workflow based on the pre-assessment
outcome and BFI-10 score. The logical flow is illustrated in
Figure~\ref{fig:appflow}.

%%% Figure 4

%% Figure 5: Onboarding

\begin{figure*}
\centering
\includegraphics[width=0.8\textwidth]{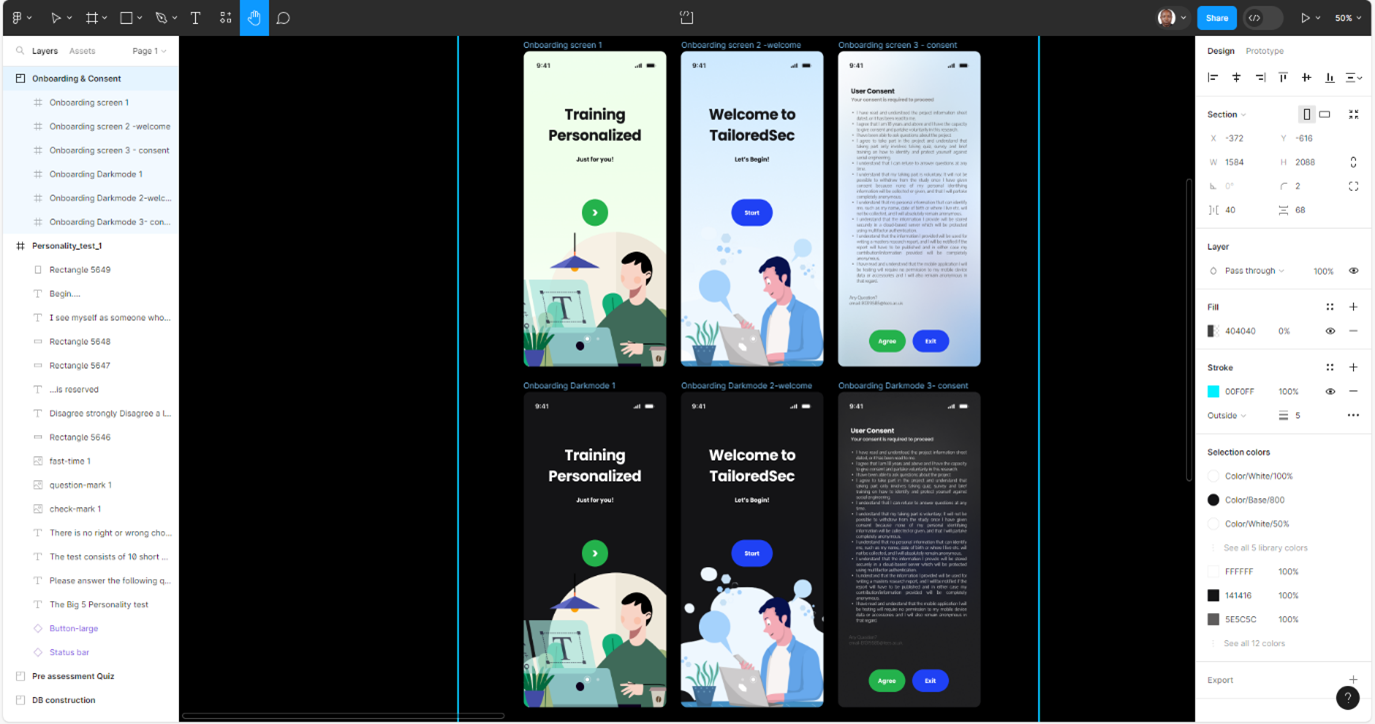}
\caption{User interface for the scenario-based pre-assessment quiz. Radio buttons
  present four response options per scenario; scoring logic is embedded in the
  navigation action of the ``Next'' button.}
\label{fig:quiz-ui}
\end{figure*}

\subsection{Onboarding and Pre-Assessment}

\begin{figure}
\centering
\includegraphics[width=0.48\textwidth]{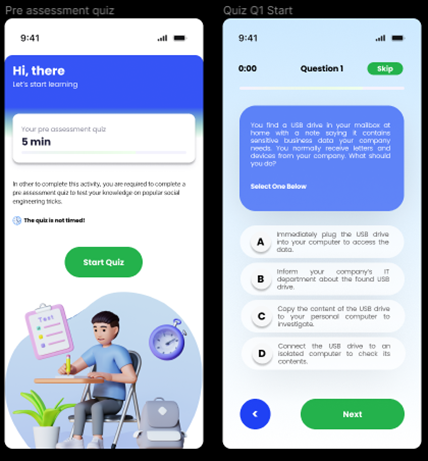}
\caption{User interface for the scenario-based pre-assessment quiz. Radio buttons
  present four response options per scenario; scoring logic is embedded in the
  navigation action of the ``Next'' button.}
\label{fig:6}
\end{figure}

The onboarding screen presents (Figure~\ref{fig:quiz-ui}) the consent statement and allows participants to select a
light or dark colour theme; no personally identifying information is requested or
collected. Upon confirming consent, participants proceed to the four-item scenario-based (social engineering scenarios) pre-assessment. Each scenario is presented as a text vignette with four radio-button
response options. The scoring logic is implemented as a conditional expression in
the mobile app state: a correct response increments the running score variable by 10
points; an incorrect response increments it by 0. Upon completion of the fourth item, the
application evaluates whether the accumulated score meets the pass threshold ($\geq 30$).
If so, the user is directed to a pass confirmation screen; if not, they are directed to a failure
screen with routing instructions. Representative screenshots of the pre-assessment
interface are shown in Figure~\ref{fig:6}.

%%% Figure 6

\begin{figure}
\centering
\includegraphics[width=0.48\textwidth]{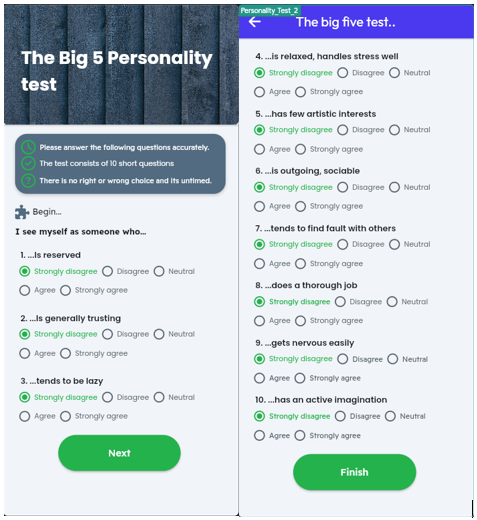}
\caption{The Big 5 Personality Test using BFI-10 model.}
\label{fig:7}
\end{figure}

\subsection{BFI-10 Scoring}
\label{subsec:bfi10scoring}

The BFI-10 was presented as ten Likert-scale items. The scoring algorithm follows
Rammstedt and John~\cite{rammstedt2007a}: items 1, 3, 4, 5, and 7 are reverse-scored
(score $\leftarrow 6 - $ response), and each trait is calculated as the sum of its two
items (see Table~\ref{tab:bfi10-scoring}). Reverse scoring was implemented by storing the
complement value in the \emph{TailoredSec} App's  integer variable at the point of user
selection. For example, a response of ``Strongly disagree'' (scale value 1) to item 1
stores the value 5 in the \texttt{persTestOpt1} state variable, yielding the correct
reverse-scored contribution to the Extraversion subscale. The dominant trait was
determined by comparing the five computed trait scores, and in the event of a tie, the
priority order $O > A > E > C > N$ was applied. All five trait scores and the resulting
dominant trait label were persisted to the database (i.e., Firebase Firestore) for the personality-conditional
group.

\begin{table}[H]
\caption{BFI-10 scoring algorithm showing item assignment and reverse-scoring
  (R = reversed). Adapted from Rammstedt and John~\protect\cite{rammstedt2007a}.}
\label{tab:bfi10-scoring}
\footnotesize
\begin{tabular}{@{}llll@{}}
\toprule
\textbf{Trait} & \textbf{Item/Score 1} & \textbf{Item/Score 2} & \textbf{Total Score} \\
\midrule
\rowcolor[HTML]{DEEAF6}
Extraversion      & 1 (R) & 6   & $Score 1 + Score 2$ \\
Agreeableness     & 2     & 7 (R) & $Score 1 + Score 2$ \\
\rowcolor[HTML]{DEEAF6}
Conscientiousness & 3 (R) & 8   & $Score 1 + Score 2$\\
Neuroticism       & 4 (R) & 9   & $Score 1 + Score 2$\\
\rowcolor[HTML]{DEEAF6}
Openness          & 5 (R) & 10  & $Score 1 + Score 2$ \\
\bottomrule
\end{tabular}
\end{table}

\subsection{Training Modules}
\label{subsec:modules}

Four personality-specific training modules were implemented, informed by the coping
mechanism recommendations of Uebelacker and Quiel~\cite{uebelacker2014social}. The
specific implementations represent pragmatic approximations to the ideal formats given
the scope and resources of the project. Table~\ref{tab:training-type} provides a
comparison of the recommended and implemented formats.

\begin{table}[H]
\caption{Recommended training formats per FFM trait compared with the implementation
  in \emph{TailoredSec}.}
\label{tab:training-type}
\footnotesize
\begin{tabular}{@{}p{1.8cm}p{2.2cm}p{3.1cm}@{}}
\toprule
\textbf{Personality} & \textbf{Recommended format} & \textbf{Implemented format} \\
\midrule
\rowcolor[HTML]{DEEAF6}
Openness          & Interactive edutainment / gamification & Swipeable flashcards (4 cards) \\
Conscientiousness & General security awareness & Awareness video \\
\rowcolor[HTML]{DEEAF6}
Extraversion      & Reward-based learning & Audio podcast + simulated loyalty-points reward \\
Agreeableness     & Narrative / storytelling & Security incident storytelling video \\
\rowcolor[HTML]{DEEAF6}
Neuroticism       & General security awareness & Awareness video (shared with Conscientiousness) \\
\bottomrule
\end{tabular}
\end{table}

%%%%%%%

\begin{figure*}[!t]
  \centering
  \setlength{\abovecaptionskip}{4pt}
  \setlength{\belowcaptionskip}{0pt}

  \begin{subfigure}[b]{0.24\textwidth}
    \centering
    \includegraphics[width=\linewidth,keepaspectratio]{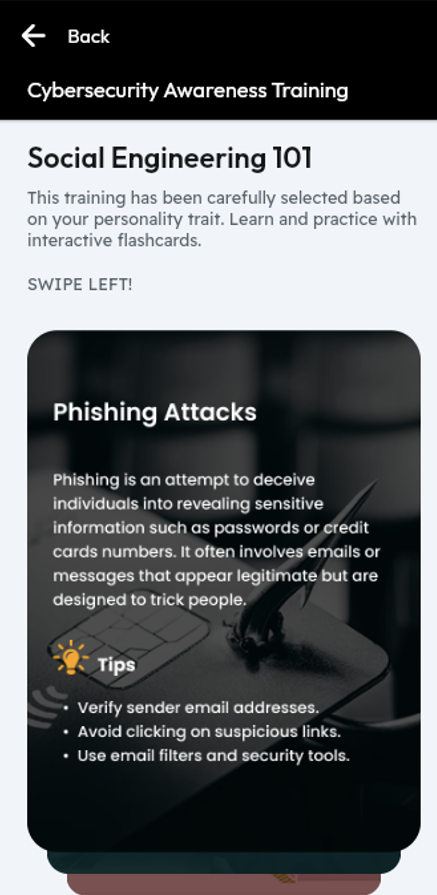}
    \caption{Swipeable Flashcards.}
    \label{fig:panel_a}
  \end{subfigure}%
  \hfill
  \begin{subfigure}[b]{0.24\textwidth}
    \centering
    \includegraphics[width=\linewidth,keepaspectratio]{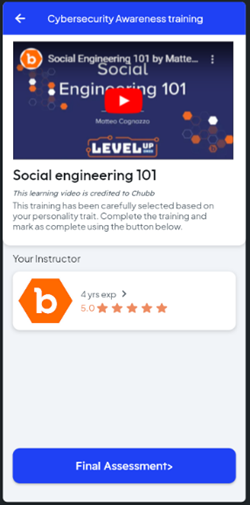}
    \caption{General Training.}
    \label{fig:panel_b}
  \end{subfigure}%
  \hfill
  \begin{subfigure}[b]{0.24\textwidth}
    \centering
    \includegraphics[width=\linewidth,keepaspectratio]{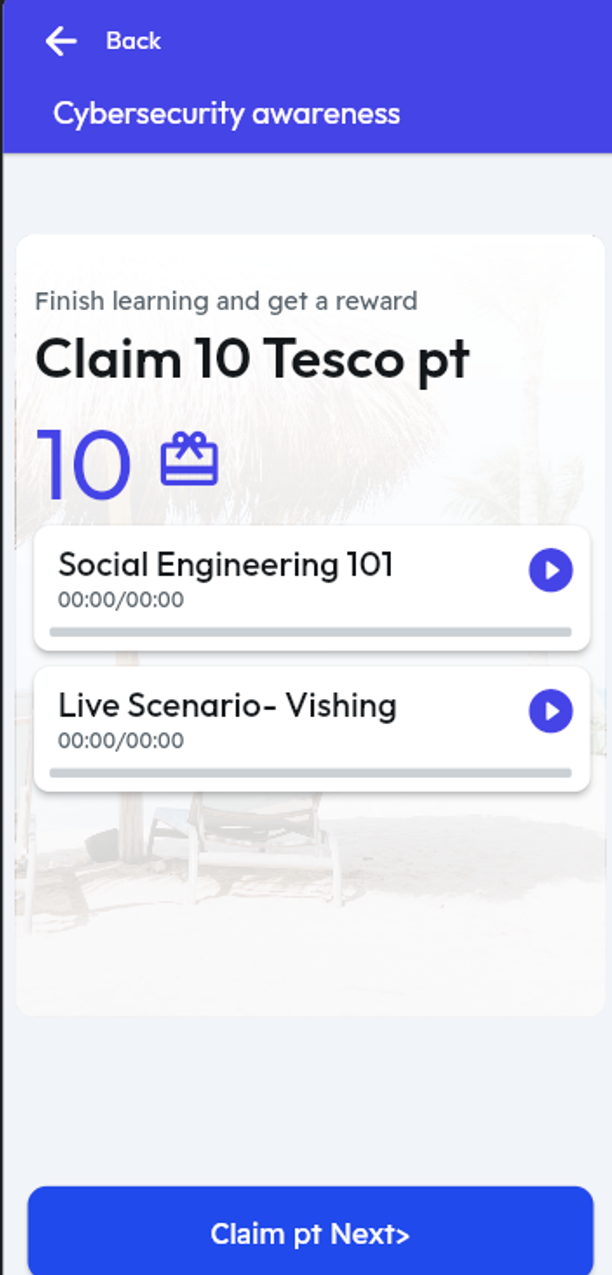}
    \caption{Reward-based Training.}
    \label{fig:panel_c}
  \end{subfigure}%
  \hfill
  \begin{subfigure}[b]{0.24\textwidth}
    \centering
    \includegraphics[width=\linewidth,keepaspectratio]{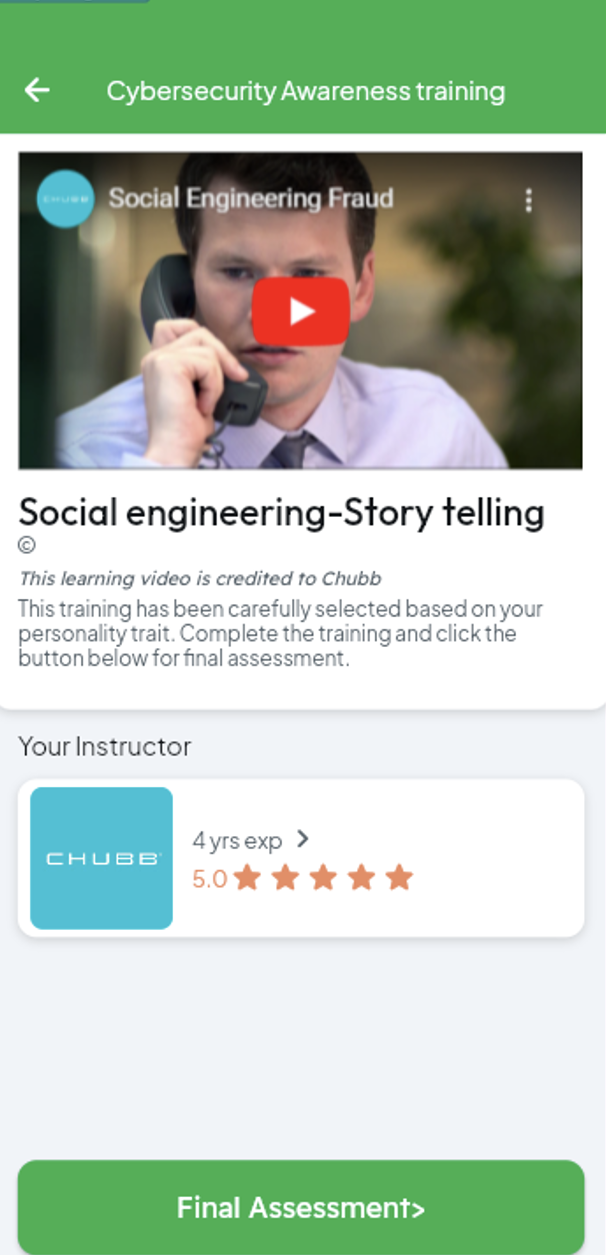}
    \caption{Story Telling.}
    \label{fig:panel_d}
  \end{subfigure}

  \caption{Trait-based Personalised Training Modules.}
  \label{fig:four_panels}
\end{figure*}
\noindent
\textbf{Openness module:} Four swipeable flashcards (Figure~\ref{fig:four_panels} (a)) cover phishing, pretexting,
tailgating, and impersonation tactics, along with detection tips. A ``Continue to assessment''
button appeared only after all four cards had been swiped, enforced by a conditional
visibility rule on the app \texttt{swipecard} counter variable.

\textbf{Conscientiousness/Neuroticism module:} A single general cybersecurity awareness
video (Figure~\ref{fig:four_panels} (b)) covered foundational social engineering concepts, policy norms, and protective
behaviours relevant to authority-based and general manipulation.

\textbf{Extraversion module:} Two audio podcasts introduced social engineering
fundamentals and presented a live vishing scenario. A simulated Tesco loyalty-point
reward (Figure~\ref{fig:four_panels} (c)) was displayed upon completion. \textit{Note:} the differential reward mechanism in
this module constitutes a potential confound, as it provides an extrinsic incentive not
present in other modules; this is acknowledged as a limitation in
Section~\ref{sec:limitations}.

\textbf{Agreeableness module:} A storytelling video (Figure~\ref{fig:four_panels} (d)) presented social engineering attacks
as narrative case studies of real-world incidents to develop empathic threat recognition
and recalibrate the trust assumptions that underlie high-agreeableness vulnerability.

\subsection{Post-Assessment and Database Design}

The post-assessment used four scenario-based MCQ items covering the same social
engineering categories as the pre-assessment but with distinct scenarios to reduce item
familiarity effects. Scoring logic was identical to the pre-assessment: correct responses
contributed 10 points to the \texttt{PostAssScore} AppState variable. Upon completion, all
session data---including pre-score, post-score, and (for the personality-conditional
group) the five BFI-10 subscale scores and dominant trait label---were recorded in a secure database (Firebase). The Firebase database was protected by server-side security rules
that restricted write access to the active session and read access to authenticated research
team credentials only.

\subsection{Technical Performance Testing}

The compiled Android APK was benchmarked using Apptim~\cite{ahmad2023analysis}, an
industry-standard mobile application profiling tool, to verify that the application
performed adequately on representative consumer hardware. The standard Apptim threshold
configuration (\texttt{example\_thresholds.yml}) was applied, with a heavy-usage
threshold set at 600\% (4200\,MB) for all resource dimensions (App memory, App CPU,
Energy score, Device CPU, Device memory, and Thread count). The application passed the
benchmark on the first run with zero crashes. Peak memory usage reached approximately
3800\,MB at application launch (approximately 15 seconds into the test) before declining
to light-usage levels. Device memory stabilised at slightly above the medium-usage mark.
All metrics remained below the heavy-usage threshold throughout the test session,
confirming that the application performs acceptably on typical consumer Android hardware.
The APK was also compiled and debugged in Android Studio with no errors.

%% ---------------------------------------------------------------

\begin{table}[H]
\caption{Pre-assessment score frequency distribution by group. Pass mark = 30/40.}
\label{tab:pre-assess}
\footnotesize
\begin{tabular}{@{}lcccc@{}}
\toprule
\textbf{Score} & \multicolumn{2}{c}{\textbf{Sample 1 (Traditional, $n=40$)}} &
                 \multicolumn{2}{c}{\textbf{Sample 2 (PC, $n=33$)}} \\
\cmidrule(lr){2-3}\cmidrule(lr){4-5}
 & $f$ & \% & $f$ & \% \\
\midrule
0  & 6  & 15.0 & 4  & 12.1 \\
10 & 11 & 27.5 & 8  & 24.2 \\
20 & 17 & 42.5 & 16 & 48.5 \\
30 & 5  & 12.5 & 4  & 12.1 \\
40 & 1  &  2.5 & 1  &  3.0 \\
\midrule
Mean (SD) & \multicolumn{2}{c}{16.00 (9.82)} & \multicolumn{2}{c}{16.97 (9.51)} \\
\bottomrule
\end{tabular}
\end{table}

\section{Results}
\label{sec:results}

\subsection{Participant Flow and Attrition}
\label{sec:results:flow}

Of 112 individuals approached, 74 completed the study (66.1\% completion rate). Sample~1
(traditional training) comprised $n = 40$ participants, and Sample~2 (personality-conditional
training) comprised $n = 34$ participants. The participant flow is shown in
Figure~\ref{fig:consort}. All 74 completers provided complete pre- and post-assessment
data. The pre-assessment frequency table for Sample~2 contains 33 entries (Table~\ref{tab:pre-assess} rather than 34, as one participant passed the pre-assessment ($\geq 30$ marks) and
proceeded directly to the post-assessment without requiring the BFI-10 routing step.) The
$t$-test analysis used $n = 34$ for Sample~2 in alignment with the post-assessment
records, which represent the complete analytical dataset.

\subsection{Baseline Equivalence}
\label{sec:results:baseline}

Pre-assessment score distributions for both groups are summarised in
Table~\ref{tab:pre-assess}. The distributions are visually similar: the modal response
in both groups was a score of 20 (Sample~1: 42.5\%, Sample~2: 48.5\%), and fewer than
15\% of participants in either group achieved a passing score ($\geq 30$) at baseline.
Group means were $M_{\text{Trad}} = 16.00$ ($SD = 9.82$, $n = 40$) and
$M_{\text{PC}} = 16.97$ ($SD = 9.51$, $n = 33$). Welch's independent-samples
$t$-test confirmed no significant difference between groups at baseline
($t(69.1) = 0.43$, $p = .67$, Cohen's $d = 0.10$), supporting the assumption of
baseline equivalence and providing a valid basis for between group comparison on the
post-assessment.

\subsection{Primary Outcome: Post-Assessment Performance}
\label{sec:results:primary}

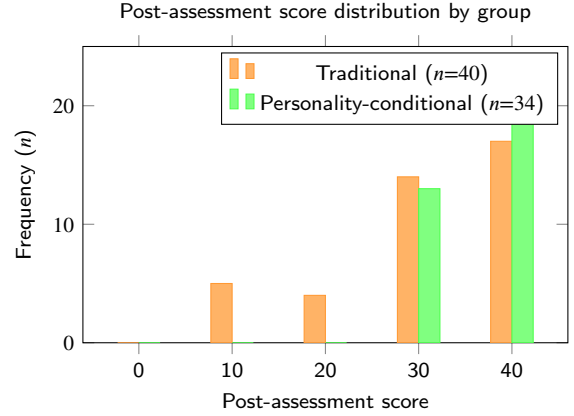
\begin{figure}
\centering
\begin{tikzpicture}
\begin{axis}[
  width=0.46\textwidth,
  height=5.5cm,
  ybar=0pt,
  bar width=8pt,
  legend style={at={(0.98,0.98)}, anchor=north east, font=\footnotesize},
  xtick={0,10,20,30,40},
  xlabel={Post-assessment score},
  ylabel={Frequency ($n$)},
  ymin=0, ymax=25,
  title={\footnotesize Post-assessment score distribution by group},
  title style={font=\footnotesize},
  label style={font=\footnotesize},
  tick label style={font=\footnotesize},
  enlarge x limits=0.15,
  symbolic x coords={0,10,20,30,40},
]
\addplot[fill=orange!60, draw=orange!80] coordinates {
  (0,0) (10,5) (20,4) (30,14) (40,17)
};
\addplot[fill=green!50, draw=green!70] coordinates {
  (0,0) (10,0) (20,0) (30,13) (40,20)
};
\legend{Traditional ($n{=}40$), Personality-conditional ($n{=}34$)}
\end{axis}
\end{tikzpicture}
\caption{Post-assessment score distributions. The personality-conditional group showed
  a highly concentrated distribution at the upper end (30--40), indicative of a ceiling
  effect in the four-item post-assessment instrument.}
\label{fig:postscores}
\end{figure}

Post-assessment score distributions are presented in Table~\ref{tab:post-assess}.
A pronounced difference in distributional shape was observed between groups: Sample~1
exhibited a broad distribution across all score levels, whereas Sample~2 showed a
highly concentrated distribution at the upper end of the scale (30 or 40), with no
participant scoring below 30. This distributional asymmetry is illustrated in
Figure~\ref{fig:postscores}.

\begin{table}
\caption{Post-assessment score frequency distribution by group.}
\label{tab:post-assess}
\footnotesize
\begin{tabular}{@{}lcccc@{}}
\toprule
\textbf{Score} & \multicolumn{2}{c}{\textbf{Sample 1 (Traditional, $n=40$)}} &
                 \multicolumn{2}{c}{\textbf{Sample 2 (PC, $n=34$)}} \\
\cmidrule(lr){2-3}\cmidrule(lr){4-5}
 & $f$ & \% & $f$ & \% \\
\midrule
0  & 0  &  0.0 & 0  &  0.0 \\
10 & 5  & 12.5 & 0  &  0.0 \\
20 & 4  & 10.0 & 0  &  0.0 \\
30 & 14 & 35.0 & 13 & 39.4 \\
40 & 17 & 42.5 & 20 & 60.6 \\
\midrule
Mean (SD) & \multicolumn{2}{c}{30.75 (10.23)} & \multicolumn{2}{c}{35.88 (5.00)} \\
\bottomrule
\end{tabular}
\end{table}

Welch's independent-samples $t$-test indicated that the personality-conditional group
scored significantly higher than the traditional group on the post-assessment
($t(58.5) = -2.81$, $p = .003$ one-tailed, $p = .007$ two-tailed). The mean difference
was 5.13 ($35.88-30.75$) marks (personality-conditional minus traditional), with a 95\% confidence
interval of $[1.47, 8.79]$ marks. The effect size was in the medium range (Cohen's
$d = 0.62$; pooled $SD = 8.25$). These results are presented in Table~\ref{tab:ttest}.

\begin{table}[H]
\caption{Welch's two-sample $t$-test on post-assessment scores (two-sample assuming
  unequal variances). PC = personality-conditional.}
\label{tab:ttest}
\footnotesize
\begin{tabular}{@{}lcc@{}}
\toprule
\textbf{Statistic} & \textbf{Sample 1 (Traditional)} & \textbf{Sample 2 (PC)} \\
\midrule
$n$         & 40          & 34            \\
Mean        & 30.75       & 35.88         \\
$SD$        & 10.23       & 5.00          \\
Variance    & 104.55      & 24.96         \\
\midrule
\multicolumn{3}{l}{Welch's $t(58.5) = -2.81$} \\
\multicolumn{3}{l}{$p$ (one-tailed) $= .003$} \\
\multicolumn{3}{l}{$p$ (two-tailed) $= .007$} \\
\multicolumn{3}{l}{Mean difference $= 5.13$ (95\,\% CI: $[1.47, 8.79]$)} \\
\multicolumn{3}{l}{Cohen's $d = 0.62$ (medium effect)} \\
\multicolumn{3}{l}{Variance ratio $= 4.19$ (ceiling effect indicator)} \\
\bottomrule
\end{tabular}
\end{table}

\subsection{Pass-Rate Analysis}
\label{sec:results:passrate}

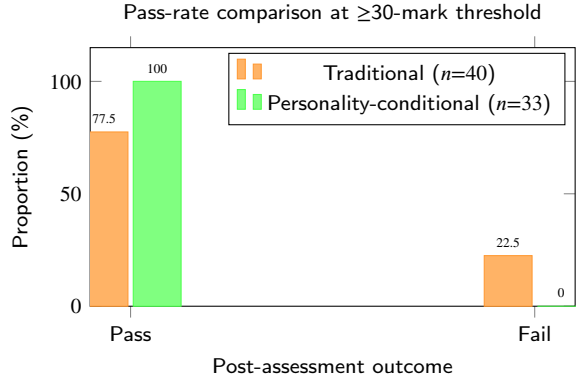
\begin{figure}
\centering
\begin{tikzpicture}
\begin{axis}[
  width=0.46\textwidth,
  height=5.0cm,
  ybar,
  bar width=18pt,
  legend style={at={(0.98,0.98)}, anchor=north east, font=\footnotesize},
  xtick=data,
  symbolic x coords={Pass, Fail},
  xlabel={Post-assessment outcome},
  ylabel={Proportion (\%)},
  ymin=0, ymax=115,
  title={\footnotesize Pass-rate comparison at $\geq$30-mark threshold},
  title style={font=\footnotesize},
  label style={font=\footnotesize},
  tick label style={font=\footnotesize},
  nodes near coords,
  nodes near coords style={font=\tiny},
]
\addplot[fill=orange!60, draw=orange!80] coordinates {(Pass,77.5) (Fail,22.5)};
\addplot[fill=green!50, draw=green!70]   coordinates {(Pass,100)  (Fail,0)};
\legend{Traditional ($n{=}40$), Personality-conditional ($n{=}33$)}
\end{axis}
\end{tikzpicture}
\caption{Pass-rate comparison at the 30-mark threshold. All 33 personality-conditional
  participants who received trait-routed training passed the post-assessment (100\%),
  compared with 31/40 (77.5\%) in the traditional group (Fisher's exact $p < .01$,
  one-tailed).}
\label{fig:passrate}
\end{figure}

The pass rate at the 30-mark threshold was 100\% (33/33) for the personality-conditional
group and 77.5\% (31/40) for the traditional group, a difference of 22.5 percentage
points. Because one cell of the $2 \times 2$ contingency table contained an expected
count of zero, Fisher's exact test was used in place of Pearson's $\chi^2$. The
one-tailed Fisher's exact test indicated a statistically significant difference in
pass-rates between groups ($p < .01$). This result is visualised in
Figure~\ref{fig:passrate}.

\subsection{User Feedback}
\label{sec:results:feedback}

A total of 68 participants completed the optional post-training usability survey.
Table~\ref{tab:feedback} summarises ratings across the four survey dimensions. Usability
and ease of use received the highest ratings: 85.3\% of respondents rated usability at
4 or 5 (``High'' or ``Highest''), and 94.1\% rated ease of use at 4 or 5. Ratings for
the quality of adaptive content matching and the app's contribution to social engineering
understanding were lower but still positive: 63.2\% rated adaptive content quality at
4 or 5, and 63.2\% rated social engineering understanding at 4 or 5. Usability satisfaction was high; the ratings for adaptive content depth are the
clearest signal for what to improve next.

\begin{table}[H]
\caption{User feedback summary ($n = 68$). Ratings on a 5-point Likert scale
  (5 = Highest, 1 = Lowest). $\geq$4 = proportion rating 4 or 5.}
\label{tab:feedback}
\footnotesize
\begin{tabular}{@{}lcccccc@{}}
\toprule
\textbf{Dimension} & \textbf{5} & \textbf{4} & \textbf{3} & \textbf{2} & \textbf{1} & \textbf{$\geq$4\,\%} \\
\midrule
\rowcolor[HTML]{DEEAF6}
Usability          & 41 & 17 &  7 & 2 & 1 & 85.3 \\
Adaptive content   & 20 & 23 & 18 & 4 & 3 & 63.2 \\
\rowcolor[HTML]{DEEAF6}
SE understanding   & 18 & 25 & 19 & 4 & 2 & 63.2 \\
Ease of use        & 47 & 17 &  4 & 0 & 0 & 94.1 \\
\bottomrule
\end{tabular}
\end{table}

%% ---------------------------------------------------------------
\section{Discussion}
\label{sec:discussion}

\subsection{Interpretation of Findings}

Routing training content by dominant BFI-10 trait produced a statistically significant
post-assessment advantage over uniform video training ($d = 0.62$, 95\,\%\,CI
$[1.47, 8.79]$ marks; Fisher's exact $p < .01$ on pass-rate). Because the two groups
did not differ at baseline ($t(69.1) = 0.43$, $p = .67$), this difference cannot
plausibly be attributed to pre-existing knowledge differences. The result aligns with
prior work showing that personality-differentiated instruction improves learning
outcomes~\cite{komarraju2011bigfive,blau2016face} and with the trait--content mapping
proposed by Uebelacker and Quiel~\cite{uebelacker2014social}.

A $d$ of 0.62 sits above the typical range ($d = 0.3$--$0.5$) reported for security
awareness interventions~\cite{aldawood2019reviewing}, which suggests that the trait-routing
mechanism contributes something beyond what generic training can achieve. The usability
ratings (85.3\% rating usability 4 or 5 out of 5; 94.1\% on ease of use) confirm that
mobile delivery was acceptable to participants and did not suppress learning itself.

\subsection{The Ceiling Effect and Its Implications}

The variance ratio of 4.19 ($s^2_{\text{Trad}} = 104.55$ vs $s^2_{\text{PC}} = 24.96$)
is the most technically significant feature of the results. It reflects a ceiling effect
in the personality-conditional group: every participant scored either 30 or 40, so the
four-item instrument had no discriminating power at all within that group above the
30-mark threshold. The true size of the post-assessment advantage is therefore
understated. A harder instrument would spread the personality-conditional scores and
would likely yield a larger effect estimate. Equally, the ceiling prevents any
distinction between participants who gained shallow familiarity and those who achieved
robust understanding. This is a strong case for using a validated, multi-item instrument
(such as the HAIS-Q~\cite{parsons2017development}) in any replication.

The same ceiling partly explains the large pass-rate increase in Sample~1 (15\% to
77.5\% across a single general video session). That jump could reflect genuine learning
from the video. It could equally reflect a testing effect, where completing the
pre-assessment primed participants for the post-assessment scenarios. Because the two
sets of items were deliberately non-parallel, the current data cannot separate the two.

\subsection{Psychometric Limitations of BFI-10 for Individual Routing}

Using BFI-10 scores to route individual participants raises a well-known psychometric
problem. Rammstedt and John~\cite{rammstedt2007a} explicitly validated the instrument
for \emph{group-level} research and warned against individual-level decisions because
the two-item per-trait structure yields retest reliabilities of only $r = .49$
(Neuroticism) to $r = .79$ (Extraversion). Taking the argmax of five subscale scores,
each derived from two items, will route a non-trivial fraction of participants to the
wrong module~\cite{rammstedt2014limits}. The present study cannot estimate how large
that fraction is; doing so would require test--retest data or a concurrent BFI-44
comparison. Future studies should either supplement BFI-10 routing with a longer
instrument in a subsample, or replace it with a more reliable measure such as the
BFI-44 or NEO-PI-3 from the outset.

\subsection{Comparison with Prior Work}

Related studies have touched on parts of this problem without assembling the whole. Thorp
et al.~\cite{thorp2023association} found that Openness moderated how participants
experienced VR-based cybersecurity training, but held content constant across trait
groups. Halevi et al.~\cite{halevi2013spear} showed that personality predicted
spear-phishing susceptibility without testing a trait-tailored intervention. The present
study is the first to implement and empirically evaluate a complete
personality-routing training system in a mobile cybersecurity application. The $d = 0.62$
effect and 100\% pass-rate in the personality-conditional group are encouraging, though
the methodological limitations catalogued here mean that any strong causal claim would
be premature.

\subsection{Practical Implications for Organisations}

For practitioners, the results are promising but warrant caution. A simple personality
screening step---ten questionnaire items before training begins---was sufficient to
produce a measurable knowledge gain over a generic video. The BFI-10's individual-level
reliability means that routing departments or teams collectively (rather than person by
person) may be the more defensible near-term approach; group-level personality profiles
are more stable than individual scores from a two-item subscale. The mobile format
itself added no apparent friction: 94.1\% of respondents rated it easy to use, and the
anonymous, no-authentication design removes common staff privacy concerns that can
suppress training uptake~\cite{isoaho2021cybersecurity}.

%% ---------------------------------------------------------------
\section{Limitations and Future Work}
\label{sec:limitations}

Eight limitations constrain the conclusions of this study.

\textbf{Assignment mechanism:} Participants were not randomly assigned to conditions.
Although baseline equivalence was confirmed on pre-assessment scores, non-random
allocation means that unobserved confounders (e.g., motivation, IT literacy, prior
cybersecurity exposure, or risk tolerance) could differ between groups. Future studies should employ
stratified random assignment, ideally with pre-registration.

\textbf{Sample size and power:} The achieved sample of $n = 74$ was below the
pre-specified target of 100. A power analysis for a one-tailed $t$-test at $\alpha = .05$
and $d = 0.62$ suggests approximately 90\% power at $n = 76$ total, so the study is
adequately powered for the observed effect, but replication with larger samples is needed
to obtain stable effect size estimates.

\textbf{Outcome instrument:} The four-item author-developed MCQ is not validated,
contains non-parallel pre/post items, and exhibits a ceiling effect after effective
training. Future studies should use a validated multi-item cybersecurity awareness
scale such as the HAIS-Q~\cite{parsons2017development} or the Security Behaviour
Intentions Scale (SeBIS), which would provide better measurement fidelity and enable
meaningful subgroup comparisons.

\textbf{Ceiling effect:} The concentration of personality-conditional post-scores at 30
and 40 marks indicates that the post-assessment was too easy for participants who received
effective training, making it impossible to detect fine-grained differences in knowledge
acquisition. A more sensitive instrument with items at multiple difficulty levels is
essential for future evaluations.

\textbf{BFI-10 psychometric limitations:} Individual-level routing based on BFI-10
argmax scores is psychometrically problematic given the instrument's two-item-per-trait 
structure and moderate retest reliability~\cite{rammstedt2007a,rammstedt2014limits}.
Future research should compare BFI-10 and BFI-44 routing decisions in the same sample
to quantify the misassignment rate or use the NEO-PI-3 for higher measurement reliability.

\textbf{Differential incentive in the Extraversion module:} The simulated loyalty-point
reward offered to participants in the Extraversion training module introduces a
differential extrinsic incentive that may have inflated engagement and post-assessment
performance for participants routed to this module, confounding trait-specific effects.
Future studies should equate incentive structures across modules.

\textbf{Selection bias from sideloading attrition:} The 33.9\% attrition is almost
entirely attributable to the requirement to sideload an unsigned APK, which
preferentially selects technically confident participants. The completer sample is
therefore not representative of the general population, limiting generalisability.
Deployment through an official app store or a web-based platform would address this bias.

\textbf{Single session and no retention follow-up:} The study used a single training
session with an immediate post-assessment, providing no information about knowledge
retention over time. A follow-up assessment at four weeks or three months is essential
to evaluate whether personality-conditional training produces durable knowledge gains.

\textbf{Missing ANCOVA:} A more rigorous analysis would use ANCOVA with the pre-assessment
score as a covariate, which provides better control for baseline individual differences
than a post-score-only comparison, even when groups are baseline-equivalent. This
analysis was precluded here by the non-parallel nature of the pre/post instruments
(different scenarios) but should be implemented in future studies using parallel forms.

Specific future work priorities, in order of impact, are: (1) a pre-registered
randomised replication with stratified allocation by personality profile and $n \geq 150$;
(2) adoption of the HAIS-Q as the primary outcome measure; (3) addition of BFI-44 for
reliability comparison with BFI-10 routing decisions; (4) a four-week knowledge
retention follow-up; (5) equalisation of incentive structures across training modules;
(6) deployment through an official app store to eliminate sideloading selection bias;
(7) screen-level engagement logging from Firebase Analytics to enable fidelity cheques;
and (8) trait $\times$ condition interaction analysis to test whether specific trait--module
pairings drive the overall effect.

%% ---------------------------------------------------------------
\section{Conclusions}
\label{sec:conclusion}

This paper presents the design, implementation, and quasi-experimental evaluation of
\emph{TailoredSec}, a mobile cybersecurity awareness application that routes
training content based on users' dominant Five-Factor Model personality trait as measured
by the BFI-10. Across a sample of 74 UK-based adults allocated to a personality-conditional
($n = 34$) or traditional ($n = 40$) training condition, the personality-conditional
group achieved a significantly higher post-assessment score (Welch's $t(58.5) = -2.81$,
$p = .003$, Cohen's $d = 0.62$, 95\,\%\,CI $[1.47, 8.79]$ marks) and a higher
pass-rate (100\% vs 77.5\%; Fisher's exact $p < .01$), despite no significant pre-assessment
difference between groups ($t(69.1) = 0.43$, $p = .67$). The application was rated
highly usable by 85.3\% of 68 feedback respondents.

These results support the idea that routing training content by personality traits is
feasible and worth pursuing more rigorously. The study contributes both a fully
specified proof-of-concept system---routing algorithm, content mapping, and BFI-10
scoring all documented to replication standards---and an empirical baseline for
comparison. The ceiling effect, BFI-10 routing reliability, and sideloading attrition
documented here translate directly into a concrete list of what any follow-on study
must fix.

Deploying personality-based training is becoming cheaper as smartphones become standard
corporate equipment. This study shows that even the simplest version---ten extra
questionnaire items before a single training session---produced a measurable advantage
over a generic video. Whether that advantage holds at scale, over time, and with a
validated outcome measure is the question the field now needs to answer.

%% ---------------------------------------------------------------
\section*{Declaration of Competing Interest}

The authors declare that they have no known competing financial interests or personal
relationships that could have appeared to influence the work reported in this paper.

\section*{Data Availability Statement}

The anonymised aggregate data supporting the results reported in this paper (pre- and
post-assessment score distributions, BFI-10 subscale scores by routing group) are
available from the corresponding author on reasonable request. Individual-level data are
not publicly available to protect participant anonymity.

\section*{Acknowledgements}

The authors thank the 74 participants who voluntarily contributed their time to this
study, and acknowledge the support of the School of Computing, Engineering and Digital
Technologies at Teesside University in facilitating ethical oversight of the research.

%% ---------------------------------------------------------------
\bibliographystyle{model1-num-names}
\bibliography{ref}

%% ---------------------------------------------------------------
\appendix

\section{Consent Notification Text}
\label{app:consent}

The following text was displayed on the application's onboarding screen prior to commencement of the pre-assessment:

\begin{quote}
\textit{Dear Participant,\\
Thank you for considering participating in this survey. By completing and submitting the quiz and survey, you consent to participate voluntarily. Your responses will be kept confidential; contact details are provided in the information sheet if you have any concern. Thank you for your participation.}
\end{quote}

\section{BFI-10 Questionnaire Items}
\label{app:bfi10}

Participants rated their agreement with the following ten items on a five-point Likert
scale (1 = Disagree strongly, 5 = Agree strongly). Items marked (R) were
reverse-scored. Items and scoring algorithm from Rammstedt and John~\cite{rammstedt2007a}.

\noindent I see myself as someone who\ldots
\begin{enumerate}[noitemsep]
\item \ldots is reserved. (Extraversion, R)
\item \ldots is generally trusting. (Agreeableness)
\item \ldots tends to be lazy. (Conscientiousness, R)
\item \ldots is relaxed, handles stress well. (Neuroticism, R)
\item \ldots has few artistic interests. (Openness, R)
\item \ldots is outgoing, sociable. (Extraversion)
\item \ldots tends to find fault with others. (Agreeableness, R)
\item \ldots does a thorough job. (Conscientiousness)
\item \ldots gets nervous easily. (Neuroticism)
\item \ldots has an active imagination. (Openness)
\end{enumerate}

\section{Assessment Quiz Questions}
\label{app:quiz}

\subsection*{Pre-Assessment Scenarios (four items)}

\textbf{Q1.} You find a USB drive in your mailbox with a note stating it contains sensitive business data your company needs. What should you do?
\begin{enumerate}[label=\alph*., noitemsep]
  \item Immediately plug the USB drive into your computer to access the data
  \item \textbf{Inform your company's IT department about the found USB drive} \hfill [correct]
  \item Copy the contents to your personal computer to investigate
  \item Connect the USB drive to an isolated computer to check its contents
\end{enumerate}

\textbf{Q2.} You receive a call from your bank asking you to verify account details to prevent unauthorised access. The caller knows some of your personal information. What is the best course of action?
\begin{enumerate}[label=\alph*., noitemsep]
  \item Provide the requested information to ensure your account's security
  \item \textbf{Politely hang up and call your bank using the official number from their website} \hfill [correct]
  \item Ask the caller for their name and credentials before proceeding
  \item Request that the caller send you an official email for verification
\end{enumerate}

\textbf{Q3.} You receive an email from what appears to be a colleague urgently requesting a confidential client database. The email contains your colleague's name, title, and company signature. What should you do?
\begin{enumerate}[label=\alph*., noitemsep]
  \item Share the database immediately to meet the urgent request
  \item \textbf{Verify the request by calling your colleague on their known phone number} \hfill [correct]
  \item Request approval from your supervisor before taking any action
  \item Reply to the email asking for additional verification
\end{enumerate}

\textbf{Q4.} A stranger in the office car park identifies themselves as a new employee who forgot their access card and asks you to swipe them in. What should you do?
\begin{enumerate}[label=\alph*., noitemsep]
  \item Swipe them in as the property is heavily guarded
  \item Request to see their company ID access card before swiping them in
  \item \textbf{Politely decline and direct them to obtain access through the proper channel} \hfill [correct]
  \item Swipe them in to be helpful since they are a new employee
\end{enumerate}

\subsection*{Post-Assessment Scenarios (four items)}

\textbf{Q1.} A colleague asks you to hold the door open for them to enter a secure area without using their access card. What should you do?
\begin{enumerate}[label=\alph*., noitemsep]
  \item Hold the door open to be courteous
  \item \textbf{Politely refuse and ask them to use their access card} \hfill [correct]
  \item Allow them entry without questioning
  \item Report the incident to security
\end{enumerate}

\textbf{Q2.} You receive an email informing you that you have won a prize and need to click a link to claim it. What action should you take?
\begin{enumerate}[label=\alph*., noitemsep]
  \item Click the link to claim your prize
  \item Forward the email to colleagues for confirmation
  \item \textbf{Delete the email without clicking the link} \hfill [correct]
  \item Click the link but do not provide any personal information
\end{enumerate}

\textbf{Q3.} You find a USB drive labelled ``Employee Bonus Details'' in the office car park. What should be your immediate action?
\begin{enumerate}[label=\alph*., noitemsep]
  \item Plug it into your computer to check the contents
  \item \textbf{Report it to the IT department} \hfill [correct]
  \item Take it home and examine its contents later
  \item Give it to a colleague to check
\end{enumerate}

\textbf{Q4.} During a work conference, a person you do not know well asks about your job role, current projects, and requests some internal company information. What is the most appropriate response?
\begin{enumerate}[label=\alph*., noitemsep]
  \item \textbf{Politely decline to share internal information and report the interaction to your security team} \hfill [correct]
  \item Share the information since you are at a professional conference
  \item Share a limited amount of information to appear cooperative
  \item Ask the person for their business card before responding
\end{enumerate}

\end{document}